\documentstyle[12pt,aaspp4, epsf]{article}
\newcommand\bb[1] {   \mbox{\boldmath{$#1$}}  }

\def\gapprox{\mathrel{\vcenter{\offinterlineskip \hbox{$>$}
    \kern 0.3ex \hbox{$\sim$}}}}
\def\lapprox{\mathrel{\vcenter{\offinterlineskip \hbox{$<$}
    \kern 0.3ex \hbox{$\sim$}}}}
\def\refindent{\par\penalty-100\noindent\parskip=4pt plus1pt
               \hangindent=3pc\hangafter=1\null}

\slugcomment{Submitted to Ap.J.}
\begin{document}
\setlength{\baselineskip}{12pt}

\title{Local Hydrodynamic Stability of Accretion Disks}
\author{John F. Hawley, Steven A. Balbus, Wayne F. Winters}
\affil{Department of Astronomy, University of Virginia, \\
Charlottesville, VA 22903; jh8h, sb, ww5m@virginia.edu}
\received{10/30/98}

\begin{abstract}

We employ a variety of numerical simulations in the local shearing box
system to investigate in greater depth the local hydrodynamic stability
of Keplerian differential rotation.  In particular we explore the
relationship of Keplerian shear to the nonlinear instabilities known to
exist in simple Cartesian shear.  The Coriolis force is the source of
linear stabilization in differential rotation.  We exploit the formal
equivalence of constant angular momentum flows and simple Cartesian
shear to examine the transition from stability to nonlinear
instability.  The manifestation of nonlinear instability in shear flows
is known to be sensitive to initial perturbation and to the amount of
viscosity; marginally (linearly) stable differentially rotating flows
exhibit this same sensitivity.  Keplerian systems, however, are
completely stable; the strength of the stabilizing Coriolis force
easily overwhelms any destabilizing nonlinear effects.  In fact,
nonlinear effects speed the decay of applied turbulence by producing a
rapid cascade of energy to high wavenumbers where dissipation occurs.
Our conclusions are tested with grid resolution experiments and by
comparison with results from a code that employs an alternative
numerical algorithm.  The properties of hydrodynamic differential
rotation are contrasted with magnetohydrodynamic differential
rotation.  The kinetic stress couples to the vorticity which limits
turbulence, while the magnetic stress couples to the shear which
promotes turbulence.  Thus magnetohydrodynamic turbulence is uniquely
capable of acting as a turbulent angular momentum transport mechanism
in disks.

\end{abstract}

\section{Introduction.}

In the past few years there has been substantial progress in
understanding the origin of angular momentum transport in astrophysical
accretion disks (see the reviews by Papaloizou \& Lin 1995 and Balbus
\& Hawley 1998).  In particular, the nature of transport by
magnetohydrodynamic (MHD) turbulence has been greatly clarified.
Magnetized disks are linearly unstable to the weak field
magnetorotational instability (Balbus \& Hawley 1991).  However, in
regions surrounding the solar nebula and in protostellar disks more
generally, temperatures and densities suggest very small ionization
fractions, leading to magnetic decoupling.  The presence of turbulence
in such a disk is problematic.  For this reason, hydrodynamical studies
of disk turbulence remain of great interest.

Before the importance of magnetic fields in disk dynamics was clear,
disk turbulence was viewed as a hydrodynamical problem.  Adverse
entropy gradients (vertical stratification) and adverse angular
momentum gradients (radial stratification) each brought with them a
legacy of local instability, turbulence and enhanced transport.
Moreover, simple shear layers break down into turbulence via nonlinear
processes, even in the absence of rotation, and Couette flow experiments
show nonlinear breakdown of some Rayleigh-stable velocity profiles
(Coles 1965; Drazin \& Reid 1981).  Even if details were a bit vague,
enhanced turbulent transport via {\it some\/} hydrodynamic process seemed
more than plausible.

Convective models of disk transport are now two decades old (Cameron
1978, Lin \& Papaloizou 1980).  But convective turbulence does not, by
itself, guarantee outward angular momentum transport (Prinn 1990);
indeed, recent investigations suggests the opposite.  Ryu \& Goodman
(1992) analyzed the linear stages of convective instability, and
pointed out that it produces inward, not outward transport.  Kley,
Papaloizou \& Lin (1993) found inward transport in an axisymmetric disk
convection simulation.  Stone \& Balbus (1996) conducted a full
three-dimensional (3D) numerical simulation of the compressible Euler
equations in a local patch of Keplerian disk, found small inward
transport, and gave arguments as to why this might be
expected.  Despite the presence of vigorous convection in these
simulations, what little net transport was present was directed
radially inwards.  The time-averaged amplitude of the stress was very
small, some three to four orders of magnitude below typical values
produced by MHD turbulence in comparable simulations.  Similar results
were found by Cabot (1996) in a 3D local Navier-Stokes calculation,
when the assumed viscosity was sufficiently small.

Shear instabilities have the virtue that any resulting transport will
certainly be directed outwards, since the outwardly decreasing angular
velocity gradient would be the source of the turbulence.  An even older
idea that convection (Crawford \& Kraft 1956), high Reynolds number
shear turbulence as a source of angular momentum transports predates
modern accretion disk theory, and is explicitly invoked in Shakura \&
Sunyaev (1973).  Unfortunately, its validity has never been
demonstrated.  Unlike convective instability, which occurs when a
well-understood linear stability criterion is violated, differentially
rotating flows are linearly stable by the Rayleigh criterion.  The
oft-made conjecture is that, despite this, Keplerian disks are
nonlinearly unstable, as evinced by some Rayleigh-stable (but
decidedly non-Keplerian) Couette flows.

In principle, the nonlinear stability question of hydrodynamical
Keplerian disks could be settled by direct numerical simulation.  But
nonlinear shear instability is a 3D problem, and the critical Reynolds
number for the onset of turbulence was thought to be too high to be
attainable with a 3D numerical code.  This, however is not so (Balbus,
Hawley, \& Stone 1996; hereafter BHS).  Working with the inviscid Euler
equations, BHS evolved numerical models at a variety of resolutions,
and for a range of angular momentum distributions.  A Rayleigh-unstable
model produced rapid growth of the perturbation energy, as expected.
Simple Cartesian shear flow also produced unstable growth, due to a
nonlinear instability.  A constant angular momentum distribution also
proved to be nonlinearly unstable:  this profile is marginally stable
to linear perturbations, and BHS used simple symmetry arguments
to show that in its stability properties the system is formally
equivalent to (unstable) Cartesian shear flow.  Thus, 3D simulations
{\it can\/} reproduce the onset on nonlinear shearing instabilities where
they are known to be present.

BHS found that simple shear and constant angular momentum flows were
the {\it only\/} (unmagnetized) Rayleigh-stable systems to exhibit any
dynamical instability.  Keplerian disks simulations, in particular,
were completely stable.  BHS argued that the crucial difference between
Keplerian flow and simple shear is the presence of Coriolis forces.
The epicyclic oscillations produced by those forces are strongly
stabilizing for both linear {\it and\/} nonlinear disturbances.
Epicyclic oscillations are not present in shear layers or in constant
angular momentum rotation profiles, which were found to be the only
nonlinear unstable flows.  If the velocity profile of a disk has even
the most gentle rise in specific angular momentum with radius, its
behavior is qualitatively different from the constant angular momentum
(or simple shear) case.

At a minimum, the findings of BHS do not augur well for the existence
of hydrodynamic turbulent transport in differentially rotating disks.
The unfavorable results of the disk convection simulations, combined
with the finding that high Reynolds number shear instabilities are
easily simulated (when present), but disappear the moment rotational
effects are smoothly introduced, suggests that only MHD turbulence
offers a viable basis for Shakura-Sunyaev (1973) $\alpha$-disk models.
If hydrodynamic turbulence is present, it must be driven by some source
other than differential rotation, and generally will not transport
angular momentum (e.g., Balbus \& Hawley 1998).

In this paper we return to the local hydrodynamic simulations and
consider the hydrodynamic stability problem from several new
perspectives.  We extend the body of simulations beyond what was done
in BHS with higher resolution, and with algorithms which differ in
their diffusive properties.  In \S2 we briefly review the moment
equations developed by BHS; these form the basis for interpreting the
results of local numerical simulations.  In \S3 we review numerical
procedures used for the local simulations.  In \S4 we investigate a
number of issues:  Is there any significant effect due to numerical
resolution on BHS's conclusions regarding the stability of Keplerian
flows?  BHS speculated that the boundary between nonlinear stability
and instability (e.g., near constant angular momentum distributions)
should not be sharp, and we confirm this expectation.  Nonlinearly
unstable, but Rayleigh-stable, laboratory Couette flows are precisely
analogous to flows which find themselves at this boundary.  We next
consider the decay of the applied turbulence in the Keplerian system,
at a number of resolutions, and with two distinct numerical schemes.
Finally we compare the Reynolds and Maxwell stresses in a series of MHD
simulations, which span a full range of background angular momentum
distributions.  In \S5 we present our conclusions.

\section{Hydrodynamic Fluctuations}

We begin with a brief review of basic disk equations and the formalism
of BHS on the nature of hydrodynamic turbulence in disk flows.

Nonadvective transport in a hydrodynamical accretion disk is determined by
the Reynolds stress tensor,  
\begin{equation}\label{one}
T_{R\phi}\equiv \langle \rho u_R u_\phi\rangle
\end{equation}
where $\rho$ is the mass density, and ${\bf u}$ is the noncircular
component of the velocity ${\bf v}$, i.e., ${\bf v} = R\Omega
\bb{\hat\phi} + {\bf u}$.  The average in
equation (\ref{one}) is spatial: we assume that a volume can be found
over which the small scale variations average out, leaving $T_{R\phi}$
a smoothly varying quantity.   The phenomenological $\alpha$ prescription
of Shakura \& Sunyaev (1973) is $T_{R\phi}=\alpha P$, where $P$ is the
dynamical pressure (possibly including radiation).  

The stress $T_{R\phi}$ has several roles.  First, and most familiar,
it is the agent of angular momentum and energy transport. 
We are particularly interested in the radial dependence of
$T_{R\phi}$.  Consider the average radial angular momentum flux,
\begin{equation}\label{momflux}
\langle R \rho v_\phi u_R\rangle  \equiv R^2\Omega \langle \rho
u_R\rangle
+ R  T_{R\phi},
\end{equation}
and the radial energy flux
\begin{equation}\label {enflux}
\langle {1\over2}\rho v_\phi^2 u_R + \Phi_c \rangle =
-{1\over 2} R^2\Omega^2 \langle\rho u_R \rangle + R \Omega  T_{R\phi}.
\end{equation}
where $\Phi_c$ is the central (external) gravitational potential.  Note
that in both equations (\ref{momflux}) and (\ref{enflux}) the first
terms represent advected flux; all nonadvective flux is in the
$T_{R\phi}$ contribution of the second terms.  Outward transport
corresponds to $T_{R\phi} > 0$.  The nonadvective contributions differ
from one another only by a factor of $\Omega$ in the energy flux.  Each
of the above (net) fluxes is a simple linear combination of $\langle
u_R\rangle$ and $T_{R\phi}$ only.  The fact that no other flow
quantities appear is crucial to the formulation of classical
steady-state $\alpha$
disk theories, for it allows for a well-defined luminosity--accretion
rate relationship.

The turbulent stress must do more than regulate angular momentum
transport, however.  It is also the conduit by which free
energy is tapped to maintain the fluctuations, which produce
$T_{R\phi}$ in the first place.  This crucially important role is 
not a part of standard $\alpha$ disk theory.  It is a consequence of
{\it fluctuation\/} dynamics, not mean flow dynamics.  This may be
seen by inspecting the diagonal moments of the radial and azimuthal
$u$ equations of motion (BHS):
\begin{equation}\label{balbusr}
{\partial\ \over\partial t}  \left\langle {\rho u_R^2\over2}\right\rangle
+ {\nabla}{\cdot}\left\langle {1\over2}\rho u_R^2 {\bf u}\right\rangle =
2\Omega T_{R\phi}
-\left\langle {u_R}{\partial P\over\partial R} \right\rangle -{\rm losses}
\end{equation}
and
\begin{equation}\label{balbusaz}
{\partial\ \over\partial t}  \left\langle {\rho u_\phi^2\over2}\right\rangle
+ {\nabla}{\cdot}\left\langle {1\over2}\rho u_\phi^2 {\bf u}\right\rangle =
-{\kappa^2\over2\Omega}T_{R\phi}
- \left\langle {u_\phi\over R}{\partial P\over\partial\phi} \right\rangle
-{\rm losses}
\end{equation}
where ``losses'' refer to viscous losses.  In disks the stress tensor
couples both to the Coriolis force and to the background shear, and the
former is bigger than the latter.

Contrast this with simple shear flows.  Here,
only the shear couple is present; the stabilizing Coriolis force is
absent.  Reverting to Cartesian coordinates with background flow
velocity $V(x) {\bf {\hat e}_y }$, the dynamical $u$-moment equations
for shear flow are
\begin{equation}\label{X}
{\partial\ \over\partial t}  \left\langle {\rho u_X^2\over2}\right\rangle
+ {\nabla}{\cdot}\left\langle {1\over2}\rho u_X^2 {\bf u}\right\rangle =
- \left\langle {u_X}{\partial P\over\partial x} \right\rangle - {\rm losses}
\end{equation}
and
\begin{equation}\label{Y}
{\partial\ \over\partial t}  \left\langle {\rho u_Y^2\over2}\right\rangle
+ {\nabla}{\cdot}\left\langle {1\over2}\rho u_Y^2 {\bf u}\right\rangle =
-{dV\over dx} T_{XY} - \left\langle {u_Y}{\partial P\over\partial y}
\right\rangle
-{\rm losses}
\end{equation}
where $T_{XY}$ is the obvious analogue to $T_{R\phi}$.  

In both disk and shear flow, the shear is the source of free energy
which maintains the kinetic energy of the fluctuations.  But the
dynamical content of (\ref{X}) and (\ref{Y}), as compared
with (\ref{balbusr}) and (\ref{balbusaz}) is evidently very different.
The disk is faced with grave difficulties in keeping up both outward
transport ($T_{R\phi} > 0)$ and significant levels of $\langle
u^2\rangle$.  Whereas $2\Omega T_{R\phi}$ is a source term for $\langle
\rho u_R^2/2\rangle$ if $T_{R\phi} >0$, the $-\kappa^2/2\Omega$ term in
equation (\ref{balbusaz}) is a sink for $\langle \rho
u_\phi^2/2\rangle$.  The presence of both a source and a sink coupled
to $T_{R\phi}$ means that the $u_R$ and $u_\phi$ fluctuations cannot
grow simultaneously: one would grow only at the expense of the other,
and the implicit correlation embodied in $T_{R\phi}$ could not be
self-consistently maintained.  One could appeal to the pressure term
in equation (\ref{balbusaz}) for help, and one needs to do so in
{\it any\/} hydrodynamical disk flow where there is outward
transport.  This leads not to turbulence, whose physical origin is
vorticity entrainment in large scale shear (Tennekes \& Lumley 1972),
but to transport by spiral waves.
In shear flow there is no $T_{XY}$
sink in the corresponding equation (\ref{Y}), and hence no barrier to
maintaining both transport and fluctuation amplitude.  The nonlinear
instability (at sufficiently high Reynolds numbers) and resulting
turbulence of simple shear
flow is a matter of common experience.   The behavior of disks could
not be more different.

\section{Numerical Procedure}

Numerical simulations demonstrate the behaviors of disk and shear flows
unambiguously.  It is sufficient to work in the local shearing-periodic
box system (Hawley, Gammie \& Balbus 1995).  The background angular
velocity of the disk is taken to be a power law: $\Omega \propto
R^{-q}$.  We construct a set of local coordinates corotating with the
fluid at a fiducial radius $R_\circ$.  Equations are expanded to first
order about $R_{\circ}$, using locally Cartesian coordinates ${\bf x} =
(x,y,z) = (R-R_{\circ}, R_{\circ}(\phi-\Omega t), z)$.  ($\Omega$ is
evaluated at $R=R_\circ$ in the expression for $y$.)  Although the
local geometry is Cartesian, Coriolis and tidal forces ensure the local
dynamics is not.

The resulting hydrodynamic equations are
\begin{equation}\label{continuity}
{\partial\rho\over{\partial t}} + \nabla \cdot (\rho {\bf v}) = 0,
\end{equation}
\begin{equation}\label{euler}
{\partial {\bf v}\over{\partial t}} + {\bf v}\cdot \nabla {\bf v}
= - {1\over\rho}\nabla  P 
        - 2 {\bf\Omega} \times {\bf v}
        + 2 q \Omega^2 x {\hat{\bf x}},
\end{equation}
\begin{equation}\label{energy}
{\partial \rho \epsilon\over{\partial t}} + \nabla\cdot(\rho\epsilon {\bf v})
        + P \nabla \cdot {\bf v} = 0,
\end{equation}
where the pressure $P$ is given by the
\begin{equation}\label{eos}
P = \rho \epsilon(\gamma - 1),
\end{equation}
and the remainder of the terms have their usual meaning.  For
simplicity, the vertical component of gravity is not included.  The
shearing box is defined to be a cube with length $L\ (=1)$ on a side, and
the initial equilibrium solution is $\rho= 1$, $P = L\Omega^2$, and
${\bf v} = -q \Omega x \hat y$.

The boundary conditions in the angular ($y$) and vertical ($z$)
directions are strictly periodic.  The radial ($x$) direction, however,
is ``shearing periodic." This means that the radial faces are joined
together in such a way that they are periodic at $t = 0$ but
subsequently shear with respect to one another.  Thus, when a fluid
element moves off the outer radial boundary, it reappears at the inner
radial boundary at its appropriate sheared position, with its
angular velocity compensated for the uniform mean shear across the
box.  See Hawley, Gammie, \& Balbus (1995) for a detailed description
of these boundary conditions.

To begin a simulation, a background angular velocity gradient
($q$ value) is chosen, and
initial velocity perturbations are introduced into the flow.  Stability
is determined by whether or not these fluctuations grow in amplitude.
The simulations of BHS began with random perturbations in pressure and
velocity applied as white noise down to the grid scale.  However, such
initial conditions have the disadvantage of varying with resolution;
the initial perturbation spectrum will never be fully resolved.  For
the models computed in this paper, we use a specific initial
perturbation rather than random noise.  The initial conditions consist
of well-defined products of sine-wave perturbations of $v_y$ in all three spatial
directions, with wavelengths $L$, $L/2$, $L/3$ and $L/4$.  A linear
combination of sine waves is constructed for each direction, e.g.,
if
\begin{equation}\label{sines}
f(x) = [\sin (2\pi x +\phi_1)+\sin(4\pi x+\phi_2)+
\sin(6\pi x+\phi_3)+\sin(8\pi x+\phi_4)]
\end{equation}
where the $\phi$ terms are fixed phase differences, then the
perturbation is applied to $v_y$ as
\begin{equation}\label{perturb}
\delta v_y = A L\Omega f(x) f(y) f(z)
\end{equation}
The amplitude $A$ of the perturbation is set to some fraction of the
shearing velocity $L\Omega$, typically 10\%.  This procedure
ensures that the initial conditions will be the same for all
simulations within a comparison group, regardless of grid resolution,
and that they will be adequately resolved, even on the $32^3$ zone
grid.

Most of the simulations described in \S 4 were computed with the same
code used in BHS.  This is an implementation of the hydrodynamic
portion of the ZEUS algorithm (Stone \& Norman 1992).  
To address the possibility of
numerical artifacts affecting our findings, it has proven
useful to compare the results obtained using two very different
numerical algorithms.  To this end, we have
adapted the Virginia Hydrodynamics-1 (VH1) Piecewise Parabolic Method
(PPM) code to the three-dimensional shearing box problem.  The PPM
algorithm was developed by Colella \& Woodward (1984), and it is a
well-known, widely-used, and well-tested numerical technique for
compressible hydrodynamics.  Like ZEUS, PPM employs directional
splitting, but differs fundamentally from ZEUS in its use of a
nonlinear Riemann solver rather than finite differences to obtain
the source terms in the Euler equations.  PPM also uses piecewise
parabolic representations (third order in truncation error) for the
fundamental variables rather than the piecewise linear
functions used in ZEUS (second order).  Both schemes employ
monotonicity filters to minimize zone to zone oscillations.  VH1 uses a
Lagrangian-remap approach in which each one-dimensional sweep through the
grid is evolved using Lagrangian equations of motion, after which the
results are remapped back onto the original grid using parabolic
interpolations.  Further information about the VH1 implementation of
PPM is currently available at http://wonka.physics.ncsu.edu/pub/VH-1,
and at http://www.pbm.com/~lindahl/VH-1.html.

\section{Results}
\subsection{Flows Marginally Stable by the Rayleigh Criterion}

A constant angular momentum distribution ($q=2$) is marginally stable
to linear perturbations by the Rayleigh criterion.  BHS showed that
such a flow, which has a vanishing epicyclic frequency, is formally
equivalent in its stability properties to simple Cartesian shear.  When
$\kappa=0$ equations (\ref{balbusr}) and (\ref{balbusaz}) have the same
form as (\ref{X}) and (\ref{Y}).  This equivalence implies that
constant angular momentum flows should be subject to the same nonlinear
instabilities that disrupt shear flows.  The simulations of BHS
demonstrate this unequivocally.


It is possible to explore deeper consequences of the symmetry.  Not
only should a $q=2$ flow be formally analogous to a shear layer, a
``$q=2-\epsilon$'' Rayleigh-stable flow should be formally analogous to
a shear layer with a little bit of rotation: $d\Omega/d\ln R \gg
2\Omega$.  This can be inferred from the $R\leftrightarrow\phi$
symmetry of equations (\ref{balbusr}) and (\ref{balbusaz}).  (From the
standpoint of a source of free energy there is no problem; differential
rotation serves this role.  This is somewhat unusual, since normally the
source of free energy disappears with the onset of linear stability.)
At large Reynolds numbers, only the ratio of the coefficients of
$T_{R\phi}$ matters, and where stability is concerned, reciprocal flows
(those whose coefficient ratios are reciprocals of one another) should
have the same stability properties.  The $q=2-\epsilon$ case is
important, because some Couette flows in which the outer cylinder
dominates the rotation are found to be nonlinearly unstable, with the
onset of instability occurring near the inner cylinder (Coles 1965;
Drazin \& Reid 1981).  This breakdown is the basis of ``subcritical''
behavior, which is occasionally cited as evidence for nonlinear
instability in {\it Keplerian\/} disks (e.g., Zahn 1991).  From the
symmetry reasons stated above, however, we believe that subcritical
behavior is evidence that disks with $q=2-\epsilon$ are nonlinearly
unstable, not $q=1.5$ disks.  This is a very testable hypothesis.

We examine this  conjecture by computing models at $64^3$ and $32^3$
resolution, $1.94\le q\le 2$ in intervals of
0.01, for two different amplitudes of initial perturbations:
$\delta v_y = 0.1 (L \Omega)$ and $\delta v_y = 0.01 (L\Omega)$.  The
value of $q$ determines the strength of the linear stabilization, the
initial perturbation amplitude sets the strength of the initial
nonlinear interactions, and the grid resolution influences the amount
of stabilizing numerical damping present [``losses'' in (\ref{balbusr})
and (\ref{balbusaz})].  Together these effects will determine when the
perturbations grow and when they do not.

Figure 1 displays some of the results.  Figure 1a shows the perturbed
kinetic energy in units of $L\Omega$ for the $32^3$ resolution, large
amplitude ($\delta v_y = 0.1L\Omega$) perturbation models.  The
different $q$ models begin with the same initial perturbations of the
form (\ref{perturb}).  The kinetic energy decreases during the first
orbit, and the curves promptly separate according to angular momentum
distribution, with the smallest $q$ model decaying the most rapidly.
Only the flows with $q=2$ and 1.99 show any subsequent growth.  The
$q=1.98$, 1.97, 1.96 and 1.95 models die away; the smaller the value of
$q$, the lower the remaining kinetic energy.  Figure 1b depicts models
with the same range of $q$ and the same initial perturbations, but
computed with $64^3$ grid zones.  Again there is a short initial period
of rapid decline in perturbed kinetic energy, with the curves
separating according to $q$.  However, this decrease is smaller than
that seen in the $32^3$ zone simulations.  After about one orbit in
time, the kinetic energies grow for all but the $q=1.96$ and 1.95
model.  These models vary with time around an average that remains
close to the initial value; only the $q=1.94$ model experiences a clear
decline in perturbed kinetic energy with time.

The sensitivity of the nonlinear instability to initial perturbation
amplitudes is demonstrated with a third group of $64^3$ grid zone
models.  These are  begun with an initial perturbation amplitude of
only $\delta v_y = 0.01 L\Omega$ (Fig. 1c).  The perturbation kinetic
energy increases slightly during the first orbit, and again the curves
separate according to $q$ value.  In this case, however, only the
$q=2.0$ model shows growth; all the others die away.

Because the instability is truly nonlinear, the details of how a flow
develops depend upon the amplitude of the initial disturbance, and, to
a far greater degree than for a linear instability, the numerical
resolution.  When $\kappa^2 = (2-q)\Omega = 0$ the linear forces on the
system sum to zero, and nonlinear dynamics determine the fate of the
flow.  The simple shear flow shows that in the absence of a restoring
force the nonlinear dynamics are destabilizing.  As $\kappa^2$ slowly
increases from zero, the linear restoring force returns; the separation
of the curves by $q$ value illustrates the stabilizing effect of the
Coriolis force.  Whether or not the linear restoring force can ensure
stability in a given system depends on its strength compared to that of
the nonlinear dynamics, which, in turn, depend on the amplitude of the
initial perturbations.  The larger the perturbations, the greater the
nonlinear forces.

The subcritical behavior of $2-\epsilon$ flows seems to have its roots
in epicyclic excursions.  The mechanism of instability in planar
Couette flow is believed to be vorticity stretching in the shear
(Tennekes \& Lumley 1972).  The presence of epicyclic motion in general
is incompatible with this process.  Nearby perturbed elements execute
bound (pressure modified) epicyclic orbits around a a common angular
momentum center.  There is no indefinite stretching of local vortices,
or at least the process is far less efficient.  But the aspect ratio of
the ellipse epicycle becomes extreme as $q\rightarrow2$; in the absence
of pressure, the minor to major (radial to azimuthal) axis ratio for
displacements in the disk plane is $(1-q/2)^{1/2}$.  At some point, it
is all but impossible to distinguish the epicycle from the shearing
background, and of course the epicyclic frequency is then tiny compared
with the shearing rate.  This latter rate is the time scale for vortex
stretching, and so we expect this mechanism for turbulence to be viable
under these conditions.  The fact the formal linear epicyclic excursion
may be bound is inconsequential.  Vortex stretching and the feeding of
turbulence will proceed if there is ample time before the epicycle
closes. For $q=1.95$, the approximate threshold for nonlinear
instability found in the simulations, $\kappa=0.2|d\Omega/d\ln R|$, and
the aspect ratio quoted above is $0.16$, i.e. about 6 to 1 major to
minor axis ratio.  These are sensible values in the scenario we have
suggested: if $\kappa$ is not less than an order of magnitude smaller
than the shearing rate, or the aspect ratio is not less than 0.1 to
0.2, then the flow is strongly stabilized by Coriolis-induced epicyclic
motion.  In this case, vortices are not stretched efficiently by the
background shear: rather than monotonically increasing the distortion,
the epicyclic orbit relaxes the vortex stretching over half of 
the period.  

Numerical diffusion error represents a loss term from (\ref{balbusr})
and (\ref{balbusaz}), and this adds to the stabilizing effect by
reducing the amplitude of the perturbations.  At a given resolution,
however, numerical effects should be nearly the same from one $q$ value
to the next.  Hence a series of models that differ by only $q$ isolates
the physical effects from the numerical.  Any differences observed in
these simulations are dynamical, not numerical, in origin.

To conclude, we have observed that the growth or decay of applied
velocity perturbations depends on the resolution and the initial
perturbation amplitude for flows near the critical $\kappa^2 = 0$
limit.  This hypersensitivity, however, is shown only over a tiny range
of $q$.  Below $q=1.95$ all indications of instability are gone.  These
results are precisely what one would expect from the presence of a
nonlinear instability, and they are consistent with the observed
presence of such instabilities in shear dominated, but formally
Rayleigh-stable Couette experiments.

\subsection{The Influence of Resolution and Algorithm}

A concern in any numerical study, particularly one whose goal is to
search for a physical instability, is the effect of finite numerical
resolution.  In \S4.1 we demonstrated how various flows
could be stabilized by increasing the epicyclic frequency (through a
decrease in $q$ from the critical value of 2.0).  In some of these
cases, a transition from stability to instability occurred when
resolution was increased.  Clearly, numerical resolution does play its
anticipated role: numerical diffusion error has a stabilizing effect.
But is numerical diffusion error decisive as $q$ becomes smaller?  

BHS argued that the stability of Keplerian flow to finite perturbations
was due to physical, not numerical effects, and gave support to that
position through simulations done at three resolutions, all of which
produced similar results.  In this section we describe a series of
simulations that improve upon those previous resolution experiments.
We have evolved a series of Keplerian flows with a range of numerical
resolutions.  We begin with $32^3$ grid zones, and then increase the
number of grid zones by a factor of two in each of the three dimensions
in subsequent simulations, up to $256^3$ grid zones.  Each of these
Keplerian flows is perturbed with angular velocity fluctuations of the
form (\ref{perturb}), with an maximum initial amplitude of $\delta v_y
= 0.1 L\Omega$.

Figure 2 shows the evolution of the kinetic energy of the angular and
radial velocity perturbations, $(\rho v_x^2)/2$ (Fig. 2a) and $(\rho
\delta v_y^2)/2$ (Fig.  2b).  The initial perturbations produce radial
kinetic energies which rapidly (0.2 orbits) increase to a maximum value comparable
with the azimuthal component.  Beyond this point, the
perturbed kinetic energies drop off with time; the higher the
resolution, the less rapid the decline, although each doubling in
resolution creates a diminishing change.  All resolutions show rapid
decrease in $(\rho \delta v_y^2)/2$ from the initial value.  One
intriguing difference between the models is that the higher the
resolution, the {\it lower} the value of $(\rho\delta v_y^2)/2$ after
about 2 orbits.  Thus, far from promoting greater instability, higher
resolution is {\it reducing} the amplitude of the angular momentum
perturbations.

Why should an increase in resolution lead to more rapid damping?  This is
clearly at variance with the expected behavior if numerical diffusion
were the sole sink of perturbed kinetic energy.  As we have
emphasized, there is also a significant dynamical sink.  Equation
(\ref{balbusaz}) shows that the Reynolds stress is a loss term for
$\langle \delta v_y^2 \rangle$.  
All simulations begin with a positive Reynolds stress,
and the higher resolution simulations maintain larger values during the
initial orbit.  At each resolution,
the Reynolds stress can be integrated over time
and space to give a measure of its strength:
$\int {\kappa^2 \over 2\Omega}\, {\langle \rho v_x\,v_y\rangle}\, dt$.
These values {\it increase} monotonically with resolution, from 0.0033,
to 0.0059, 0.0078, and finally to 0.0092 for the $256^3$ model.  (For
reference, the initial value of $\langle{{1\over 2} \rho v_y^2 }\rangle$
is 0.04.)

Further evidence for the damping effect of the Reynolds stress can be
seen in the low resolution run.  In Figure 3 we plot $(\rho \delta
v_y^2)/2$ (Fig. 3a) and the Reynolds stress (Fig.  3b) as a function of
time for the first orbit in the $32^3$ grid zone model.  This low
resolution simulation is of special interest because at orbit 0.25 the
averaged Reynolds stress becomes negative.  At the same time, the rate
of decline of $\langle \rho \delta v_y^2 \rangle$ decreases, as one
would anticipate from (\ref{balbusaz}).  Hence, although at low
resolution grid scale numerical diffusion is the largest loss term in
the angular velocity fluctuation equation, the sink due to the Reynolds
stress is large enough to observe directly.  Improved numerical
resolution increases the dynamical Reynolds sink by a greater factor
than it reduces the numerical diffusion!

We next turn to simulations of Keplerian flows at $32^3$, $64^3$ and
$128^3$ grid zone resolutions using the VH1 PPM code.  We ran the same
problem with the same initial perturbations as above.  Figure 4 shows
the time-history of the perturbed radial and azimuthal kinetic
energies.  This plot should be compared with Figure 2 and for reference
we include the $32^3$ and the $128^3$ ZEUS simulation results as
dashed lines.  Figure 5 is the Reynolds stress during the first orbit
for all the resolutions and for both algorithms; the PPM runs are the
bold lines.

The PPM results are completely consistent with the ZEUS simulations.
Most striking is the close similarity between a given PPM evolution and
the ZEUS simulation run at twice its resolution.  For example, the
history curve of the Reynolds stress in the PPM $32^3$ run lies almost
on top of the ZEUS $64^3$ curve (fig. 5) through 0.2 orbits in time.
The Reynolds stresses in the $64^3$ and $128^3$ PPM simulations peak at
the same level as the $256^3$ ZEUS simulation, then decline at slightly
different rates beyond 0.2 orbits.  The $128^3$ PPM simulation
apparently has less numerical diffusion than the $256^3$ ZEUS model.
Regardless of the relative merits of the two schemes, the succession of
curves with increasing resolution showing the same outcome, done with
two completely independent algorithms, indicates convergence to a
solution near that of the maximum resolution models.  In other words,
Keplerian disks would prove stable even if computed at arbitrarily high
resolution.

\subsection{Nonlinear Decay in the Keplerian System}

In simulations of Keplerian differential rotation, the kinetic energy
of the perturbations declines at a rate which itself decreases with
time.  Why should there be any decay at all in a stable inviscid
system?   Is this decay entirely a numerical artifact?

These simulations begin with perturbations of the form
(\ref{perturb}).  The initial power spectrum for the perturbed kinetic
energy thus contains power in the first four wavenumbers only.  Once
the evolution begins, nonlinear interactions cause a cascade of
available perturbed kinetic energy into higher wavenumbers.
Dissipation occurs rapidly at the highest wavenumber.  Although this
dissipation is numerical, it mimics the behavior of physical
dissipation at the viscous scale.  The rate at which energy cascades to
larger wavenumbers, and hence the rate at which the perturbed kinetic
energy declines, should again be a function of numerical resolution and
perturbation amplitude.  In this section we investigate these effects,
explore the reasons for the decay of the turbulence, and examine the
properties of the velocity fluctuations that remain at late time.

A study the Fourier power spectrum of the perturbed kinetic energy
yields important information.  Because the background velocity has
shear, we must transform the data into coordinates in which the
shearing box system is strictly periodic, take the Fourier transform,
and then remap the wavenumbers onto the fixed Eulerian system.  This
procedure is described in Hawley et al.\ (1995).  Figure 6 is a one
dimensional power spectra, $| \delta {\bf v}(k)|^2$ in $k_x$,
$k_y$, and $k_z$, for the $64^3$ and $128^3$ grid zone Keplerian PPM
simulations discussed in \S4.2.  The spectra are shown for orbits 1, 2
and 3, with the dashed lines corresponding to the $64^3$ run and the
solid lines to the $128^3$ model.  The initial perturbation spectrum is
constant across the first four wavenumbers (seen in Figure 6 as a
horizontal line).

Immediately after the evolution begins, energy cascades into higher
wavenumbers.  Within one third of an orbit, a relatively smooth power
law distribution has been established.  As time goes by the energy at
all wavenumbers steadly declines.  The power spectrum across the
smallest wavenumbers remains relatively flat but has dropped steadily
from $t=0$.  Beyond $k\sim 10$ the spectra drop off as steep power
laws, with the $k_y$ distribution the steepest of all three
directions.  Because of the background shearing velocity, transport in
$y$ direction produces the largest numerical diffusion.  The $k_x$
function has the smallest slope.  In this case, the background shear
causes low $k_x$ waves to be wrapped into higher wavenumbers, i.e.,
$k_x(t) = k_x (0) - tm d\Omega/dR$, where $m$ is an azimuthal
wavenumber.  The higher resolution curves in Figure 6 have
larger energies compared to the low resolution curve.  Aside from this,
the additional grid zones extend the curves out to higher
wavenumber without much significant qualitative difference.

Next we consider the effect of the initial perturbation amplitude on
the rate of decay of the turbulence and the properties of the system at
late times.  Experience suggests that if a system is vulnerable to
nonlinear instabilities, large initial perturbation amplitudes will
promote the onset of turbulence.  Indeed, this is what was observed in
the marginally Rayleigh-stable runs described in \S4.1.  Here we run a
series of low resolution $32^3$ Keplerian simulations that have initial
velocity perturbations with maximum values equal to $\delta v_y/L\Omega
= 1.0$,  0.1, 0.01, and 0.001.  The time evolution of the perturbed
kinetic energies in these four runs is shown in Figure 7.  All the runs
show rapid decay; contrary to naive expectations, however, the higher
the initial perturbation amplitude the {\it larger} the initial decay
rate of the perturbed kinetic energy.

Figure 8 illustrates the reasons for this.  Figure 8 shows the 1D
Fourier power spectrum for the largest and smallest initial
perturbation runs after 0.25 orbits.   The importance of nonlinear
interactions is increased by larger initial amplitudes.  This is why in
\S4.1 nonlinear effects were able (in some cases) to overcome the
stabilizing influence of the Coriolis force when the initial
perturbation amplitudes were increased.  Here, however, the main
nonlinear effect promoted by increased perturbation amplitude is to
create a more rapid cascade of energy to high wavenumbers.  In
contrast, the $\delta v_y = 0.001L\Omega$ case is dominated by linear
and numerical effects.   Energy has been carried into higher $k_x$
wavenumbers by linear shear, and lost from $k_y$ by numerical diffusion
error.  The spectrum in $k_z$ is completely flat through the first four
wavenumbers (those initially powered) before dropping precipitously.

Evidence for a strong nonlinear cascade in the largest intial
perturbation runs is also found in the rapid increase in entropy at the
start of the evolution due to thermalization of the initial kinetic
energy in those runs.  By orbit 5, the decay rates have been reduced to
a much lower level comparable to that seen in the small amplitude
perturbation runs.  The ratio of the kinetic energy at orbit 5 to the
initial value in each of these runs is 0.00042, 0.0042, 0.014, and
0.058.  Eventually the fluctuation energy in all these simulations
levels off at a finite, small value.  What remains at these late times
are pressure and epicyclic waves, whose amplitude is determined by the
strength of the initial perturbation.  The very slow numerical decay of
these long-wavelength linear waves is due to numerical dissipation.
The Reynolds stress oscillates around zero in all runs, with an
amplitude similar to the late-time kinetic energy.

We have argued that the residual kinetic energy represents nothing more
than linear waves left over from the initial perturbations.   Their
presence does not imply that Keplerian disks are somehow still
``slightly unstable''; stability certainly does not require that
velocity perturbations die out.  Indeed, a complete decay to zero
amplitude would have been puzzling; the motivation of the section
after all was to give an account of why there was {\it any\/} decay.
This understood, even low levels of velocity fluctuations might
be of interest in a disk, if they could be sustained indefinitely.  Can
one convincingly rule out the possibility that these lingering
fluctuations are somehow feeding off the differential rotation?  An
experiment to test this conjecture is to chart the evolution of
perturbations in a $q=0$, constant $\Omega$ flow.  In a uniformly
rotating disk, Coriolis forces are present without any background shear
at all.  Such a system is rigorously stable; without background shear
there is no source of energy to feed velocity perturbations.  At late
times, the noise in a uniformly rotating disk must reflect residual
energy from the initial conditions, not ongoing excitation.  Further,
the absence of shear flow will reduce the effective numerical
viscosity; the perturbations will not be advected by the shear flow,
nor will radial wavenumber modes be sheared out to higher values.

The $q=0$ case has been run at two resolutions, $64^3$ and $32^3$, for
comparison with equivalently resolved Keplerian systems.  The initial
perturbations have a maximum value $\delta v_y = 0.1 L\Omega$.  The
time histories of the perturbed kinetic energy for both the $q=0$ and
the $q=1.5$ $64^3$ simulations are shown in Figure 9.  Both angular
velocity distributions show rapid declines in kinetic energy, although
after 10 orbits the rate of decline is greatly reduced.  The $32^3$
resolution simulations look similar, except that they have less energy
at late time.  The residual energy is due to stable waves that have not
yet been damped by numerical diffusion.  Compared to similarly resolved
simulation with a Keplerian rotation profile ($q=1.5$) the $q=0$ models
level out at {\it higher\/} energies.  Without advection through the
grid, there is less numerical diffusion, and higher residual wave
amplitudes are preserved.  The case for driven residual weak Keplerian
turbulence becomes untenable, if the ``turbulence'' is stronger in a
rigorously stable uniformly rotating disk!

\subsection{The Contrast with Magnetic Turbulence}

Although Keplerian flows have proven to be stable to the local
development of hydrodynamic turbulence, the inclusion of a magnetic
field changes everything, even if the field is weak (subthermal).
Hydrodynamic stability in a differentially rotating system is assured
so long as the Rayleigh criterion $dL/dR > 0$ is satisfied.  Magnetic
differentially rotating systems quite generally require $d\Omega/dR >
0$ for stability (Balbus 1995), a condition not satisfied in accretion
disks.  With a magnetic field present the stress tensor acquires a
magnetic component proportional to $B_R B_\phi$,
\begin{equation}\label{magstress}
T_{R\phi}= \langle \rho u_R u_\phi - \rho u_{AR}u_{A\phi}\rangle,
\end{equation}
where
\begin{equation}
{\bf u_A} = { {\bf B}\over \sqrt{4\pi\rho}}.
\end{equation}
Most importantly, the way the stress tensor couples to the
fluctuations changes.  With the new expression for $T_{R\phi}$
the mean flow equations (\ref{momflux}) and (\ref{enflux}) are
unchanged, but the fluctuation equations become
\begin{equation} \label{magenr}
{1\over2}{\partial\ \over\partial t}\langle \rho (u_R^2 +u_{A\, R}^2)\rangle
+\nabla {\cdot}\langle\quad \rangle=
2\Omega\langle\rho u_R u_\phi \rangle -
\langle u_R {\partial P_{tot} \over \partial R} \rangle - {\rm losses,}
\end{equation}
\begin{equation} \label{magenaz}
{1\over2}{\partial\ \over\partial t}\langle \rho (u_\phi^2 +u_{A\,
\phi}^2)\rangle
+\nabla{\cdot} \langle\quad \rangle =
 - 2\Omega\langle\rho u_R u_\phi \rangle
 - T_{R\phi}\,{d\Omega\over d\ln R}
 - \langle {u_\phi\over R} {\partial P_{tot} \over \partial
 \phi}\rangle -\rm{losses}.
\end{equation}
(Terms proportional to $\nabla{\cdot} {\bf u}$ have been dropped,
the fluxes are not shown explicitly, and
$ P_{tot} = P + {B^2/8\pi}$.)

Now the stress tensor no longer works at cross purposes to itself.
There is still Coriolis stabilization in equation (\ref{magenaz}), but
it is not sufficient to overcome the stress--gradient coupling term.
One consequence of this is the now well-understood linear instability
of weak magnetic fields in disks (Balbus \& Hawley 1991; see reviews by
Papaloizou \& Lin 1995, and Balbus \& Hawley 1998).  Another is that
outward transport of angular momentum maintains the turbulence
self-consistently by directly tapping into the free energy of
differential rotation.

The different couplings of the Maxwell (magnetic) and Reynolds stresses
can be demonstrated in simulations.  Abramowicz, Brandenburg, \& Lasota
(1996) carried out a series of simulations with different values of
background $q$.  They found an increase in angular momentum transport
levels roughly in proportion to the background shear to vorticity
ratio, i.e., $q/(2-q)$.  This result is best understood by rewriting
the right hand side of (\ref{magenaz}) to obtain
\begin{equation}\label{qstress}
{1\over R}{dR^2\Omega\over dR}\langle\rho u_R u_\phi\rangle
- {d\Omega\over d\ln R} \langle\rho u_{AR} u_{A\phi}\rangle.
\end{equation}
Thus the Reynolds (kinetic) stress couples directly to the vorticity
[$=(2-q)\Omega$], and the Maxwell (magnetic) stress couples to the shear
($q\Omega$).  In other words, vorticity limits turbulence whereas shear
promotes it.

Here we expand the study of Abramowicz et al.\ by examining a full range of
$q$ values between 0 and 2 in intervals of 0.1 in a local shearing
box.  The simulations are of the same type as some of those presented
in Hawley, Gammie \& Balbus (1996).  The initial magnetic field is $B_z
\propto \sin(2\pi x/L_x)$ with a maximum corresponding to $\beta =
P/P_{mag} =400$.  The box size is $L_x = L_z = 1$, and $L_y = 2\pi$,
and the grid resolution is $32\times 64\times 32$.

Figure 10 shows time-averaged Reynolds and Maxwell stresses as a
function of $q$ for the full range of simulations.  The magnetic
instability is present for all $q>0$.  Equation (\ref{qstress})
provides no direct limit on the Maxwell stress; it acquires whatever
level the nonlinear saturation of the instability can support.
However, if the turbulence is to be sustained from the differential
rotation, not pressure forces, the Maxwell stress must in general
exceed the Reynolds stress by more than $(2-q)/q$, the ratio of the
vorticity to the shear.  In practice the ratio of the Maxwell stress to
Reynolds stress is significantly greater than this, particularly in the
range $0<q<1$.  In this regime the vorticity is so strongly stabilizing
that the Reynolds stress is kept to a minimum even when fluid
turbulence is created and maintained by the magnetic instability.  When
$q>1$, however, the shear and vorticity become comparable; the Reynolds
and Maxwell stresses both climb with increasing $q$.  As $q\rightarrow
2$, the vorticity goes to zero and there are no constraints on the
Reynolds stress from (\ref{qstress}).  The total stress increases
dramatically as the flow enters the domain of the nonlinear
hydrodynamical instability.  When $q>2$, of course, the flow is
Rayleigh unstable.

\section{Discussion}

In this paper we have carried out a series of numerical simulations to
explore further the local hydrodynamical stability properties of
Keplerian disks, and the role that the Reynolds stress plays in
determining that stability.  The key conceptual points are embodied in
the moment equations (\ref{balbusr}) and (\ref{balbusaz}) for
hydrodynamics, and (\ref{magenr}) and (\ref{magenaz}) for
magnetohydrodynamics.  The differences in those equations are clearly
manifest in simulations, both hydrodynamic and MHD.  The Maxwell stress
couples to the shear, the Reynolds stress to the vorticity.  While the
former maintains turbulence, the latter works against it.  Thus, while
magnetized disks are unstable, and naturally create and sustain
turbulence, a nonmagnetized Keplerian flow possesses only the Reynolds
stress, and that cannot by itself sustain turbulence.  The accumulating
evidence, both numerical and analytic, from this paper and earlier
works (BHS; Stone \& Balbus 1996), points clearly to the conclusion
that Keplerian flows are locally hydrodynamically stable, linearly and
nonlinearly.

It has been traditional to point to the nonlinear instabilities
observed in simple shear flows to support the conjecture that Keplerian
disks behave similarly.  Such reasoning, however, neglects the critical
difference between such flows, namely the dynamical stabilization due
to the Coriolis force.  Linear stabilization is measured by the
epicyclic frequency, $\kappa^2 = 2(2-q)\Omega^2$.  As
$q\rightarrow 2$, $\kappa^2 \rightarrow 0$, and dynamical stabilization
becomes weaker and weaker.  At $q=2$ it vanishes completely; the flow
becomes equivalent to a simple Cartesian shear and subject to the
nonlinear instabilities to which simple shear flows are prone.  Viewed
in this light, the nonlinear instability of a Cartesian shear flow is
less a generic property than a singular case lying between the linearly
unstable and linearly stable regimes.  The nonlinear instability exists
not because nonlinear forces can generally overcome linear restoring
forces, but because those linear forces vanish at the delimiting
boundary between
Rayleigh stability ($q<2$) and instability ($q>2$).

This is highlighted by our study of the transition between stability
and instability.  By varying $q$ to values near to but slightly less
than 2, we can explore the dynamics of systems close to the marginal
stability limit.   We find that when stabilization from the Coriolis
term is very weak, both the amplitude of the initial perturbations and
the size of the numerical diffusion error (grid resolution) can
determine whether the velocity perturbations amplify or decay.  This is
entirely consistent with the experimental configurations that are
linearly stable but which nevertheless become unstable.  Such
nonlinearly unstable systems are precisely those where a large shear
dominates over other factors (e.g., a rapidly rotating outer cylinder
in a Couette experiment).  In these fluid experiments the transition to
turbulence depends on the amplitude of the perturbing noise and the
Reynolds number of the flow.  When we reduce the strength of the
Coriolis force  by varying $q$ just below the marginally stable value
$q=2$, we produce a similar dominance of shear and again find an
instability that depends on the initial perturbation amplitude and the
(numerical) viscosity.  We have understood this in terms of epicyclic
orbits, which are highly distorted near $q =2$, almost
indistinguishable from background shear.  Once $q$ is sufficiently
below $q=2$, however, Coriolis stabilization is powerful, epicycles are
rounder, and perturbation growth is no longer possible.

This conclusion is greatly strengthened by experiments in which the
Keplerian system is evolved with different initial perturbations and
different grid resolutions.  First we explored the impact of finite
resolution.  Recall that the effect of numerical diffusion error on
flow structure (the turbulent perturbations) will be as an additional
loss term in (\ref{balbusr}) and (\ref{balbusaz}).  Even if we were to
assume an ideal scheme with no numerical losses, however, the sink due
to the Coriolis term in (\ref{balbusaz}) would remain unchanged.  The
simulations with various $q$ values near but just below 2 provide a
quantitative measure of just how big that term needs to be to stabilize
the flow, and an estimate of the importance of numerical viscosity as a
loss term.  Although we find that increasing the effective Reynolds
number (i.e., by increasing the resolution and thusly reducing
numerical diffusion) can convert a marginally stable flow into a
marginally unstable one, one should not conclude that further increases
will have a similar effect on strongly stable Keplerian flows.  Vortex
stretching can ``sneak'' into a highly elongated epicycle, but it
cannot do so in a rounded, strongly stable Keplerian disturbance.

We have investigated the possibility of diffusive numerical
stabilization with a series of resolution experiments run with two
completely different algorithms.  Keplerian simulations were run at 4
resolutions from $32^3$ up to $256^3$ using the ZEUS hydrodynamics
scheme, and 3 resolutions from $32^3$ up to $128^3$ using the PPM
algorithm.  The results from all these simulations were very similar.
No hint of instability was seen in any of these simulations, nor was
there any trend observed which could conceivably suggest instability in
an extrapolation to arbitrarily high resolution.  Furthermore, not just
decaying trends, but detailed numerical behavior was reproduced in two
distinct codes with very different numerical diffusion properties.  The
case that numerical diffusion is dominating and stabilizing these runs
is untenable.

Next, a series of experiments explored a range of initial perturbation
amplitudes.  The largest had initial fluctuations that were comparable
to the background rotation velocity $L\Omega$.  We found that the {\it
larger} the initial perturbation, the more rapid the decay of the
resulting turbulence.  Far from promoting instability, stronger initial
perturbations actually increase the rate of decay of the perturbed
kinetic energy.  When finite amplitude perturbations are added to the
Keplerian system they rapidly establish a nonlinear cascade of energy
to higher wavenumbers.  This energy is then thermalized (or lost,
depending upon the numerical scheme and the equation of state) at the
high wavenumber end.  Linear amplitude perturbations do not decay via
such a cascade, and damp at much lower rates.

Turbulence decays in homogeneous systems lacking an external energy
source.  A uniformly rotating disk is an example of such.  A Keplerian
system is more interesting because decay is observed
despite the presence of free energy in the differential rotation which
could, in principle, sustain the turbulence.  This does not happen,
however, because the coupling of the Reynolds stress to the background
vorticity simply makes it impossible to power simultaneously both the
radial and azimuthal velocity fluctuations that make up the Reynolds
stress.  Residual levels of fluctuations were even lower in the
Keplerian disk than they were in the uniformly rotating disk.

This behavior stands in contrast to the MHD system.  The magnetic
instability produces not just turbulent fluctuations, but the {\em
right kind\/} of fluctuations:  positive correlations in $u_R$ and
$u_\phi$, and in $B_R$ and $B_\phi$.  It is because the
magnetorotational instability is driven by differential rotation that
the critical $R$--$\phi$ correlations exist.  Unless $T_{R\phi}$ were
positive, energy would not flow from the mean flow into the
fluctuations.  Hydrodynamical Cartesian shear flows maintain the
correlation physically by ensnaring vortices (a nonlinear process);
magnetic fields do this by acting like springs attached to the fluid
elements (a linear process).  Sources of turbulence other than the
differential rotation (or simple shear) do not force a correlation
between $u_R$ and $u_\phi$, and generally do not lead to enhanced
outward transport.

Magnetic fields, then, are uniquely suited to be the agents responsible
for the behavior of $\alpha$ disks.  While this conclusion has
important implications for fully ionized disks, its implications for
protostellar disks are yet more profound.  If such disks are unable to
sustain adequate levels of magnetic coupling, or unable to sustain such
coupling throughout their radial extent, angular momentum transport may
well be tiny or nonexistent.  Angular momentum transport, when it
occurs, will have to be accomplished through global nonaxisymmetric
waves, driven, for example, by self-gravitational instabilities.  Even
if triggered by, say, convective instability, turbulence would likely
prove to be transient: it cannot be sustained from the only source of
energy available, namely the differential rotation.  More generally,
nonmagnetic disks will not be describable in terms of the usual
$\alpha$ model.

Phenomenologically much less is known of MHD turbulence than of
hydrodynamical turbulence.  There is very little laboratory to draw
upon, in contrast to the rich literature of hydrodynamical Couette
flow.  The observational complexity of many disk systems suggests the
presence of a versatile and eruptive source of turbulence; magnetic
fields seem an obvious candidate for producing such behavior.  The
physics behind magnetic reconnection, large scale field topology,
ion-neutral interactions, magnetic Prandtl number, and global dynamos
is likely to prove at least as rich and complex as the behavior of
astrophysical disks.  

This work is supported in part by NASA grants NAG5-3058, NAG5-7500, and
NSF grant AST-9423187.  Simulations were carried out with support from
NSF Metacenter computing grants at the Pittsburgh Supercomputing Center
and at NPACI in San Diego.

\clearpage
\begin{center}
{\bf References}
\end{center}

\refindent Abramowicz, M., A. Brandenburg, \& J.-P. Lasota 1996,
MNRAS, 281, L21
\refindent Balbus, S.~A. 1995, ApJ, 453, 380
\refindent Balbus, S.~A., \& Hawley, J. F. 1991, ApJ, 376, 214
\refindent Balbus, S.~A., \& Hawley, J.~F. 1998, Rev Mod Phys, 70, 1
\refindent Balbus, S.A., Hawley, J.F., \& Stone, J.M. 1996, ApJ, 467,
76 (BHS)
\refindent Cabot, W. 1996, ApJ, 465, 874
\refindent Cameron, A.~G.~W. 1978, Moon and Planets, 18, 5
\refindent Colella, P., \& Woodward, P. R. 1984, J. Comput. Phys., 54, 174
\refindent Coles, D. 1965, J. Fluid Mech, 21, 385
\refindent Crawford, J.~A., \& Kraft, R.~P. 1956, ApJ 123, 44
\refindent Cuzzi, J.~N., Dobrovolskis, A.~R., \& Hogan, R.~C. 1996, in
Chondrules and the Protoplanetary Disk, ed. R. H. Hewins, R. H.
Jones, \& E. R. D. Scott (Cambridge: Cambridge Univ. Press), 35
\refindent Drazin, P. G., \& Reid, W. H. 1981, Hydrodynamical Stability
(Cambridge: Cambridge University Press)
\refindent Hawley, J.~F., Gammie, C.~F., Balbus, S.~A. 1995, ApJ, 440,
742
\refindent Hawley, J.~F., Gammie, C.~F., Balbus, S.~A. 1996, ApJ, 464,
690
\refindent Kley, W., Papaloizou, J. C. B., \& Lin, D. N. C. 1993, ApJ,
416, 679
\refindent Lin, D.~N.~C., \& Papaloizou, J.~C.~B. 1980, MNRAS, 191, 37
\refindent Papaloizou, J.~C.~B., \& Lin, D.~N.~C. 1995, ARAA, 33, 505
\refindent Prinn, R.~G. 1990, ApJ, 348, 725
\refindent Ryu, D. \& Goodman, J. 1992, ApJ 388, 438
\refindent Shakura, N.~I., \& Sunyaev, R.~A. 1973, A\&A, 24, 337
\refindent Stone, J.~M., \& Balbus, S.~A. 1996, ApJ, 464, 364
\refindent Stone, J.~M., \& Norman, M.~L. 1992, ApJS, 80, 753
\refindent Tennekes, H., \& Lumley, J. L. 1972, A First Course in
Turbulence (Cambridge:  MIT Press)
\refindent Zahn, J-P. 1991, in Structure and Emission Properties of
Accretion Disks, C. Bertout, S. Collin-Souffrin, J-P. Lasota, \& J.
Tran Thanh Van eds (Gif sur Yvette, France:  Editions Fronti\`eres)

\newpage

\begin{figure}
\plotone{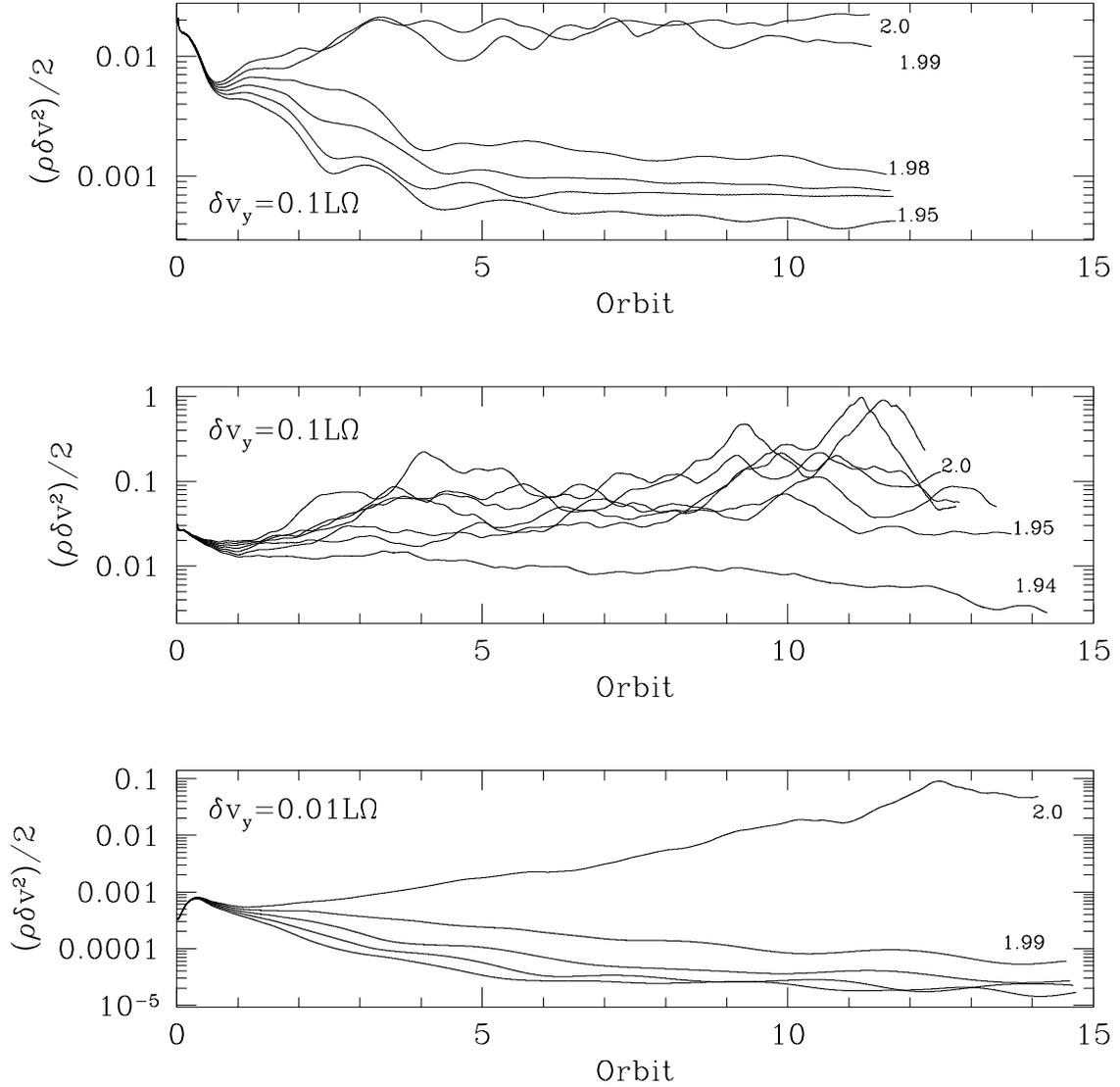}
\caption{Evolution of kinetic energy of velocity perturbations for
background rotation laws $\Omega \propto R^{-q}$ near the marginally
stable constant angular momentum distribution $q=2$.   Selected curves
are labeled by their $q$ value.  Top:  Low resolution $32^3$ grid zone
simulations with initial maximum perturbation $\delta v_y =
0.1L\Omega$.  Only the upper two curves ($q=2$ and $q=1.99$) show any
perturbation amplification.  Middle:  Simulations with $64^3$ grid zone
resolution and initial perturbation amplitude $\delta v_y =
0.1L\Omega$.  The 6 curves correspond to $q=2.0$ to $q=1.94$ in
increments of 0.01.  The $q=1.95$ curve remains level while the
$q=1.94$ declines with time.  Bottom:  Simulations with $64^3$ grid
zones and initial perturbation amplitude $\delta v_y = 0.01L\Omega$.
There 5 curves range from $q=2.0$ to $q=1.96$ in increments
of 0.01.  Only the $q=2.0$ curve shows growth.
}
\end{figure}

\begin{figure}
\plotone{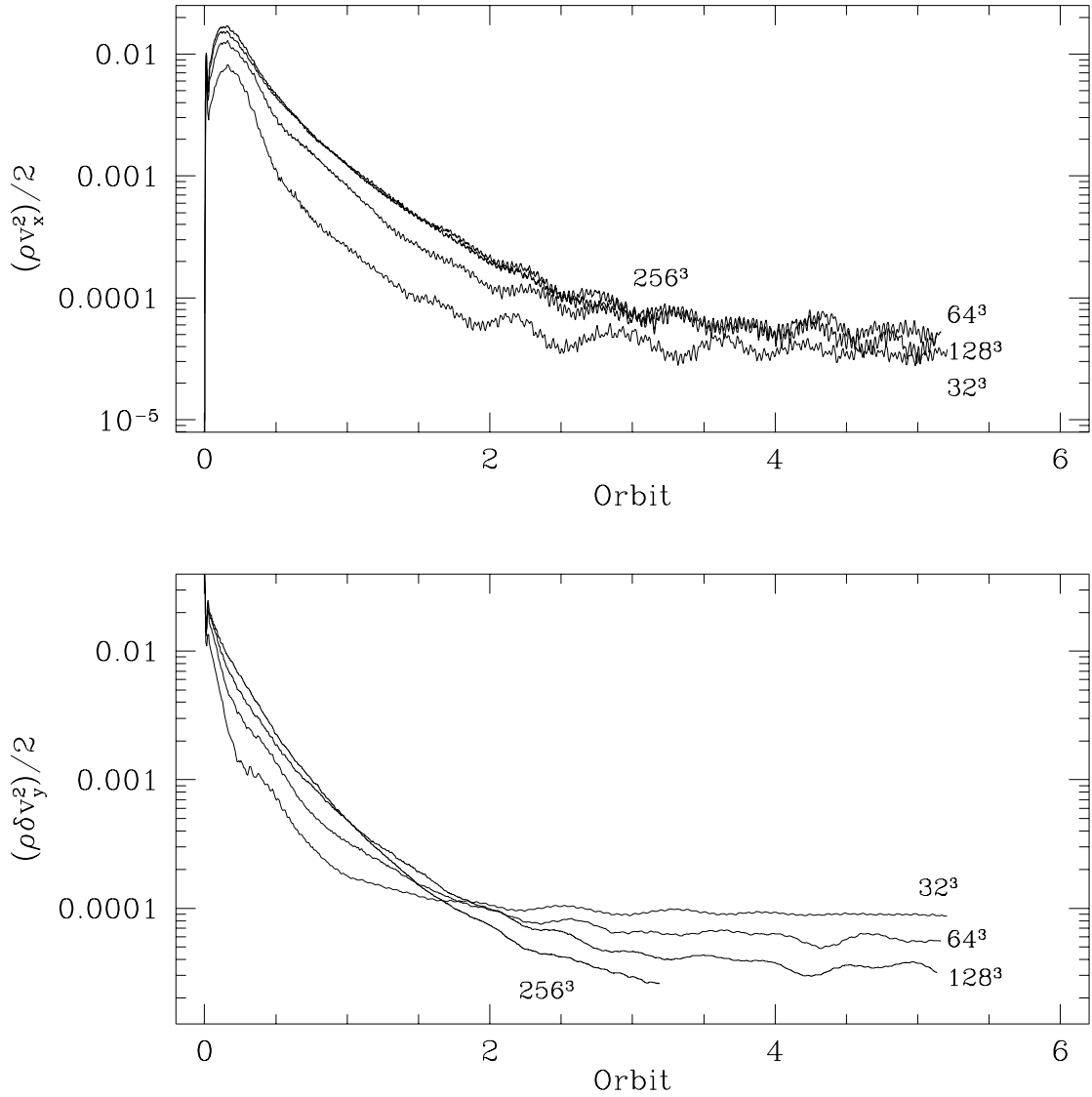}
\caption{
Evolution of $v_x$ (top) and $v_y$ (bottom) fluctuation
kinetic energy for simulations with resolutions of $32^2$, $64^3$,
$128^3$, and $256^3$ grid zones.  Curves are labeled by resolution.
}
\end{figure}

\begin{figure}
\plotone{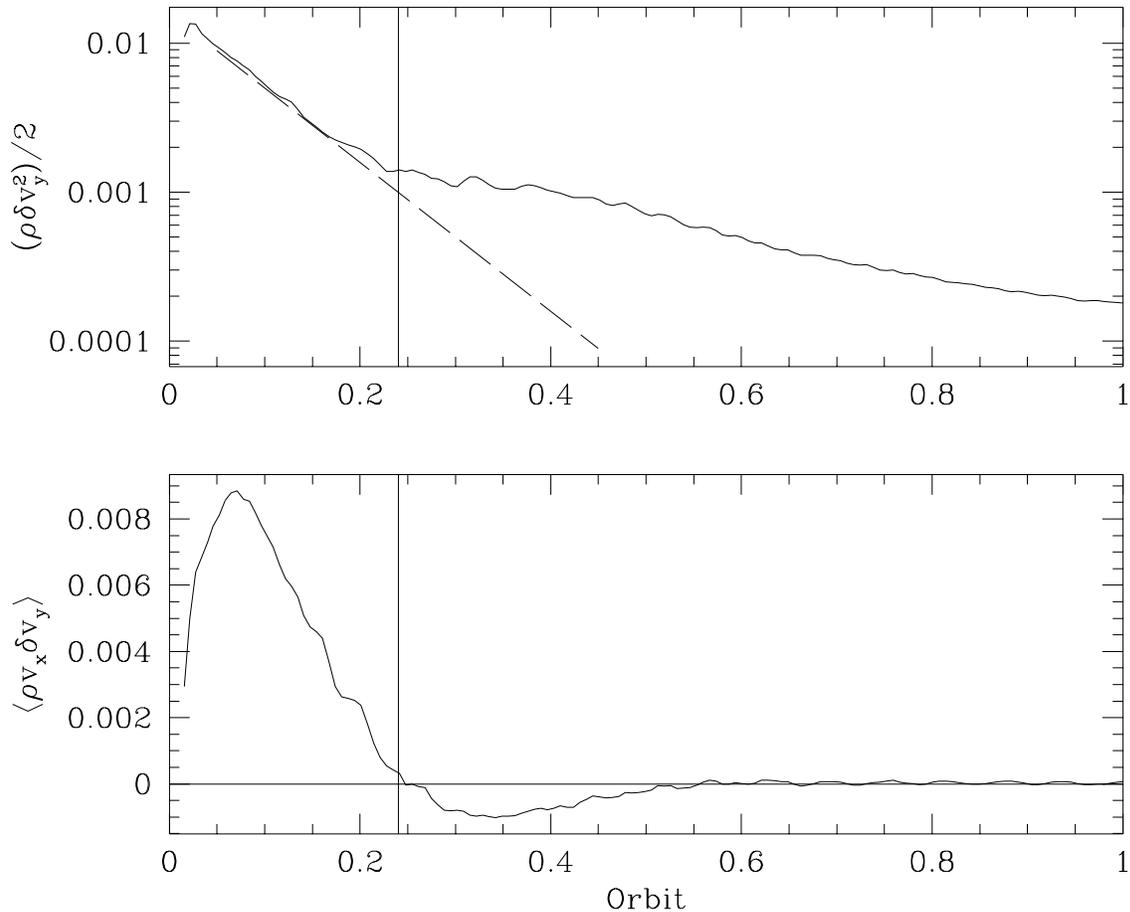}
\caption{
Time evolution of perturbed angular velocity kinetic energy
and volume-averaged Reynolds stress for the $32^3$ simulation.  The
abrupt change of slope in $\rho\delta v_y^2/2$ (dashed line added for
reference) that occurs at $t=0.24$ (indicated by vertical line)
corresponds to the point in time when the Reynolds stress becomes
negative.  A negative Reynolds stress provides a source for angular
velocity fluctuation energy; a positive Reynolds stress is a sink.
}
\end{figure}

\begin{figure}
\plotone{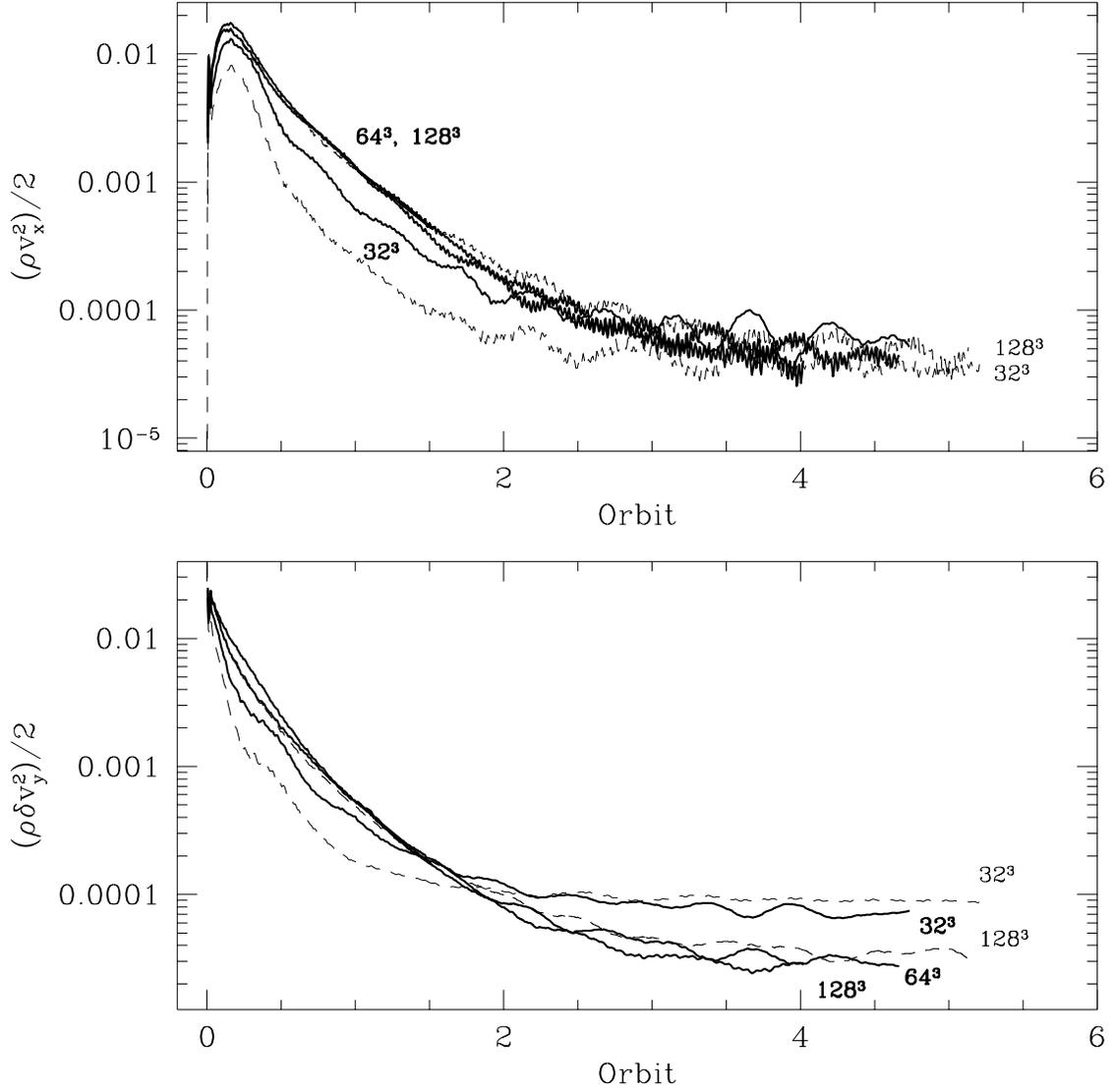}
\caption{
Evolution of $v_x$ (top) and $v_y$ (bottom) fluctuation
kinetic energy for 3 simulations using the PPM algorithm with $32^2$,
$64^3$, and $128^3$ grid zones (bold curves).  The $32^3$ and $128^3$
grid zone simulations from Figure 2 (dashed curves) are included for
reference.
}
\end{figure}

\begin{figure}
\plotone{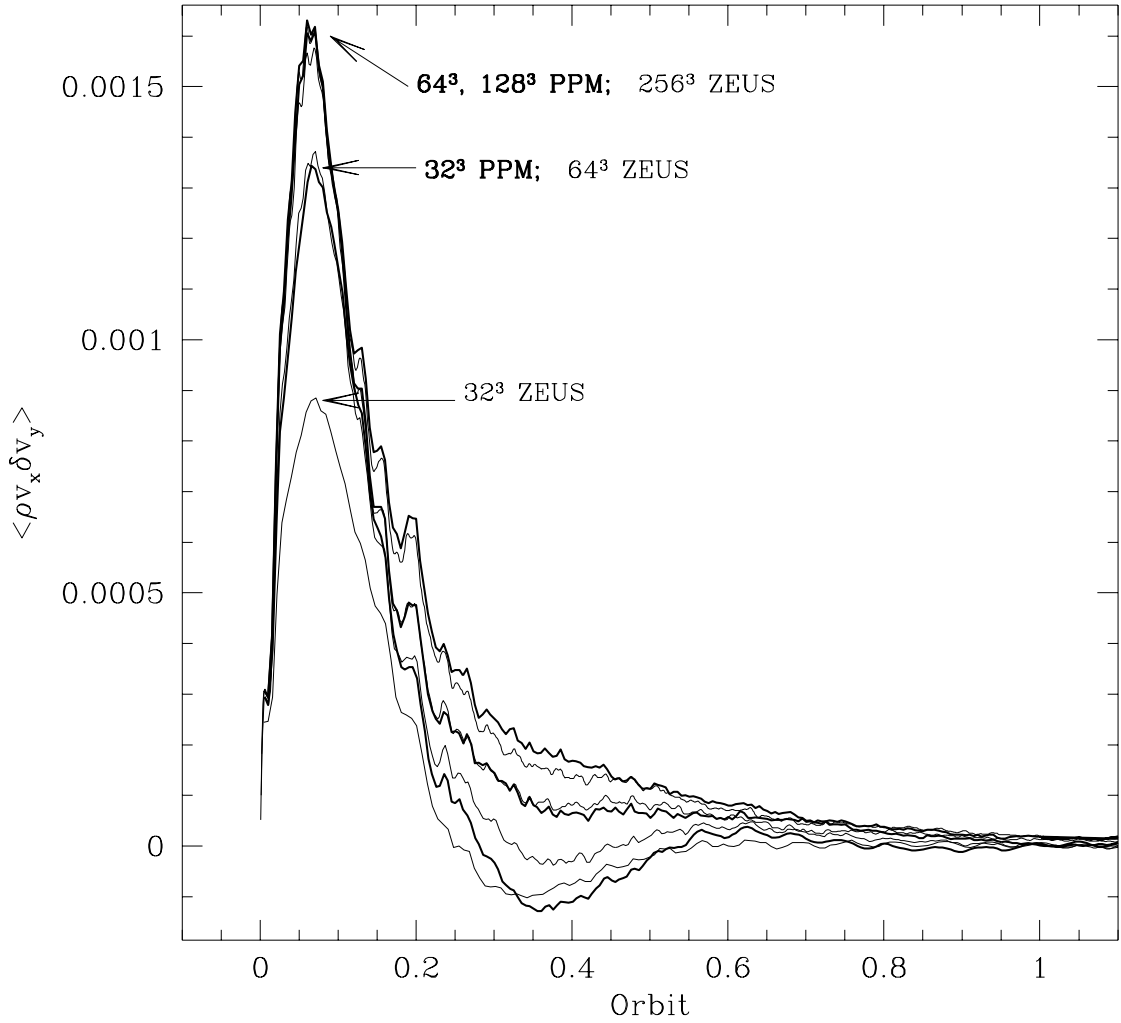}
\caption{
Time evolution of the Reynolds stress over the first orbit
in the Keplerian simulations for a range of resolutions and for both
the ZEUS and PPM (bold lines) numerical algorithms.   The peak in the
stress is labeled by the corresponding resolution and algorithm.
}
\end{figure}

\begin{figure}
\plotone{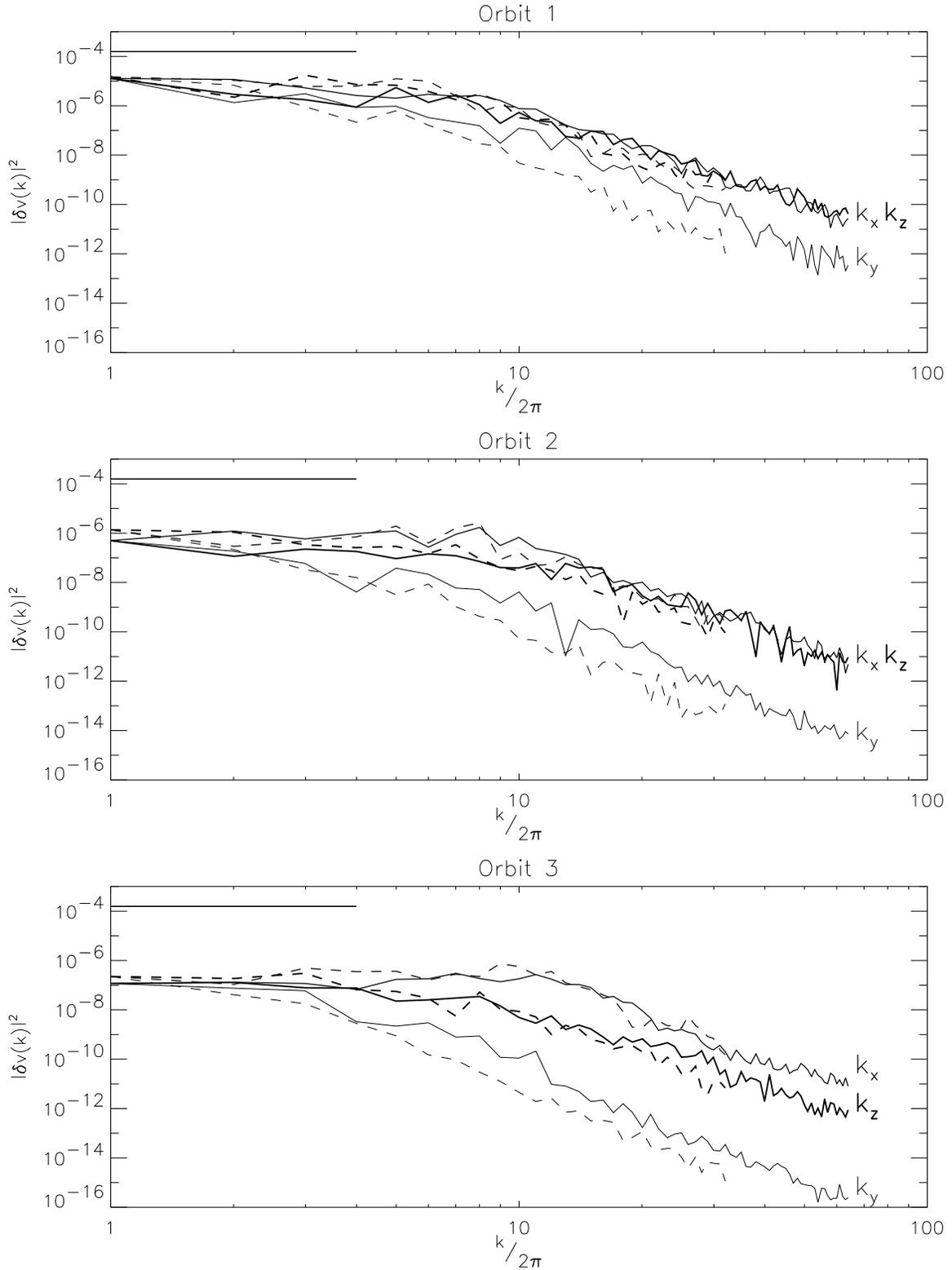}
\caption{
One dimensional power spectrum $|\delta v(k)|^2$
for the $128^3$ (solid line) and $64^3$ (dashed line) PPM Keplerian
simulation at 1, 2 and 3 orbits.  The horizontal line extending out to
$k/2\pi = 4$ is power spectrum of the initial perturbation. 
}
\end{figure}

\begin{figure}
\plotone{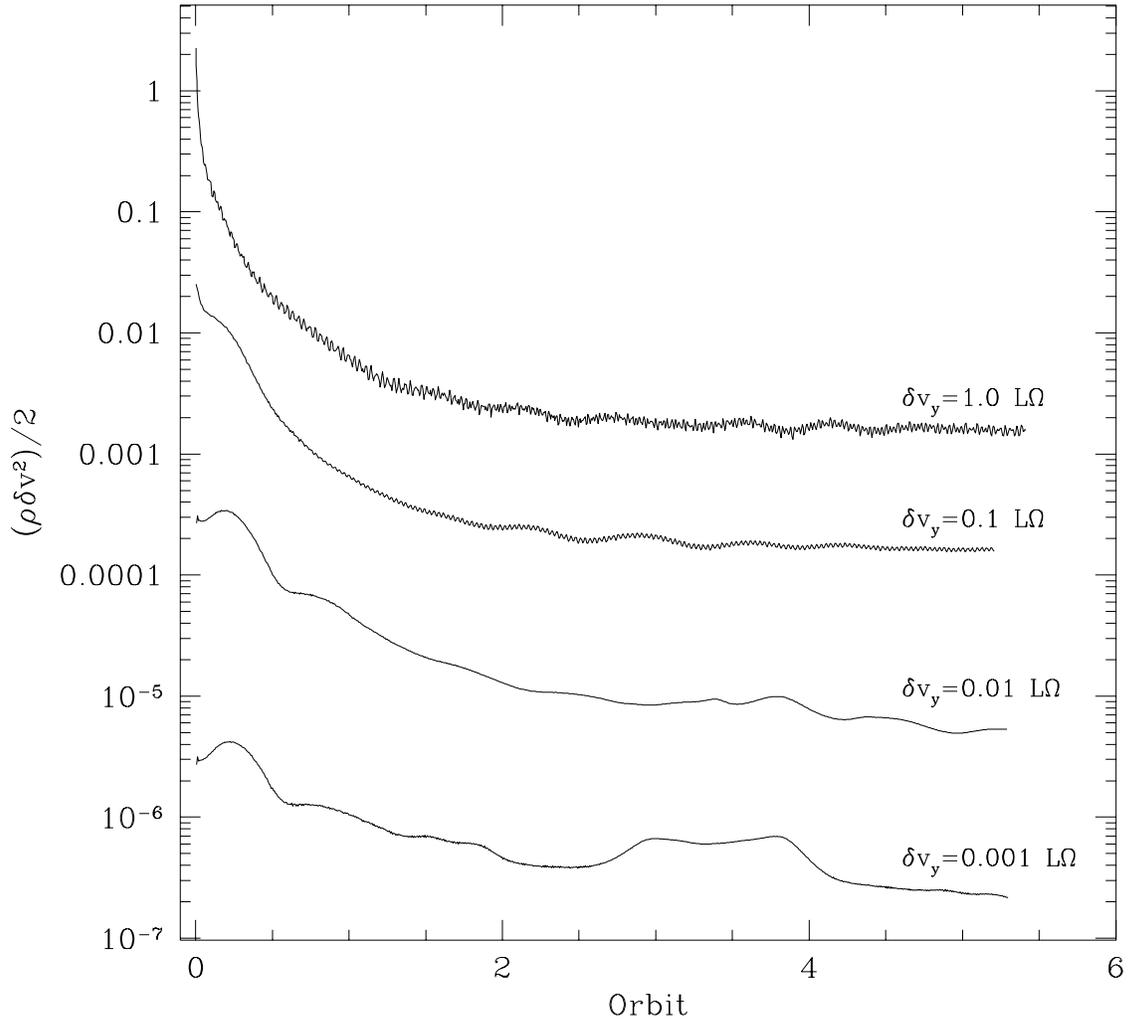}
\caption{
Time evolution of the perturbation kinetic energy in four
$32^3$ grid zone simulations of Keplerian shearing systems.  The curves
are labeled by the maximum amplitude of the initial perturbations.
Larger initial perturbations show a larger rate of decay of the
perturbed kinetic energy.
}
\end{figure}

\begin{figure}
\plotone{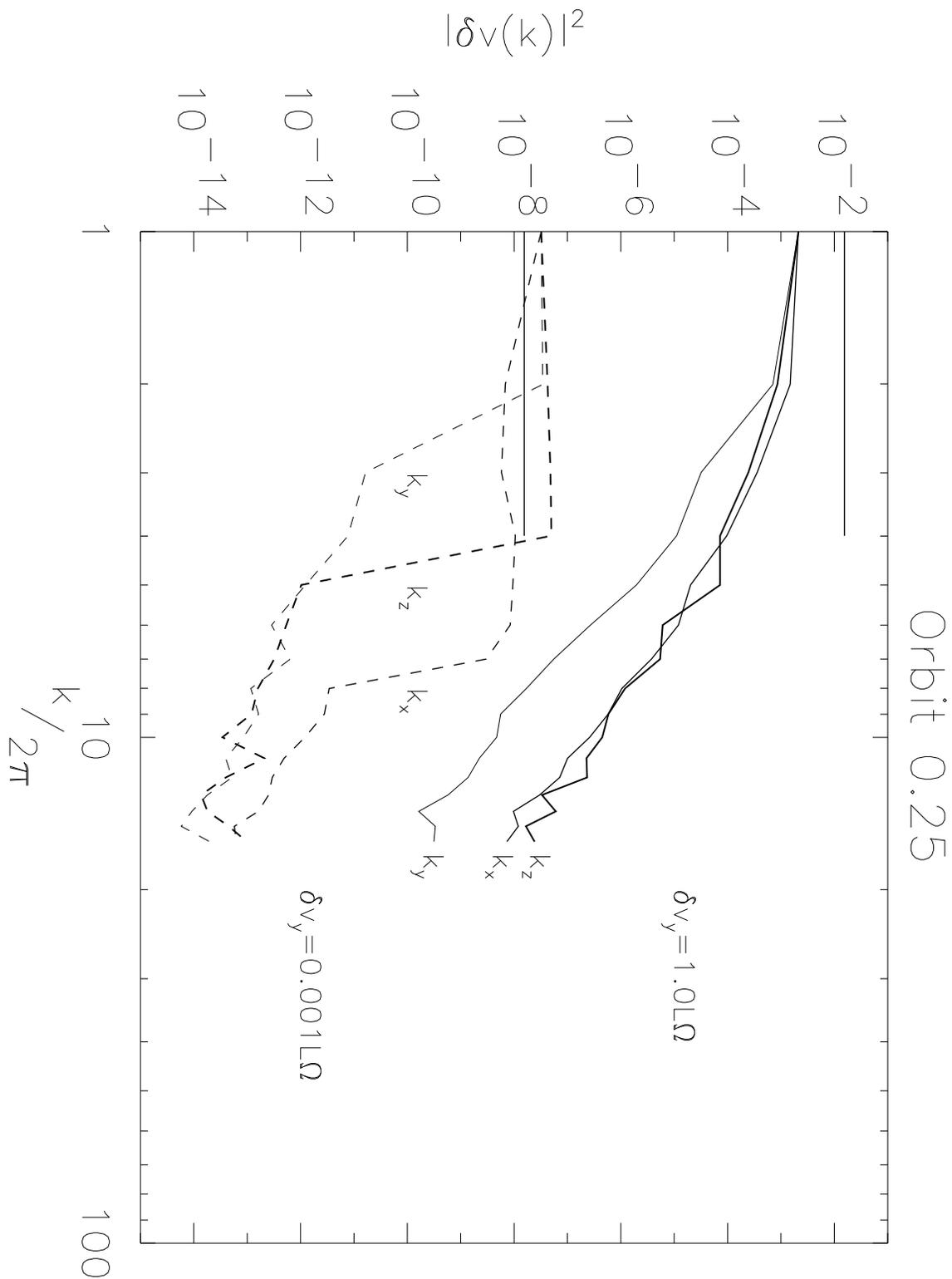}
\caption{
One dimensional power spectrum $|\delta v(k)|^2$ at 0.25
orbits for a $32^3$ large amplitude perturbation simulation (solid line) 
and a $32^3$ small amplitude perturbation simulation (dashed line).
The curves are labeled by their initial maximum perturbation amplitude.
}
\end{figure}

\begin{figure}\plotone{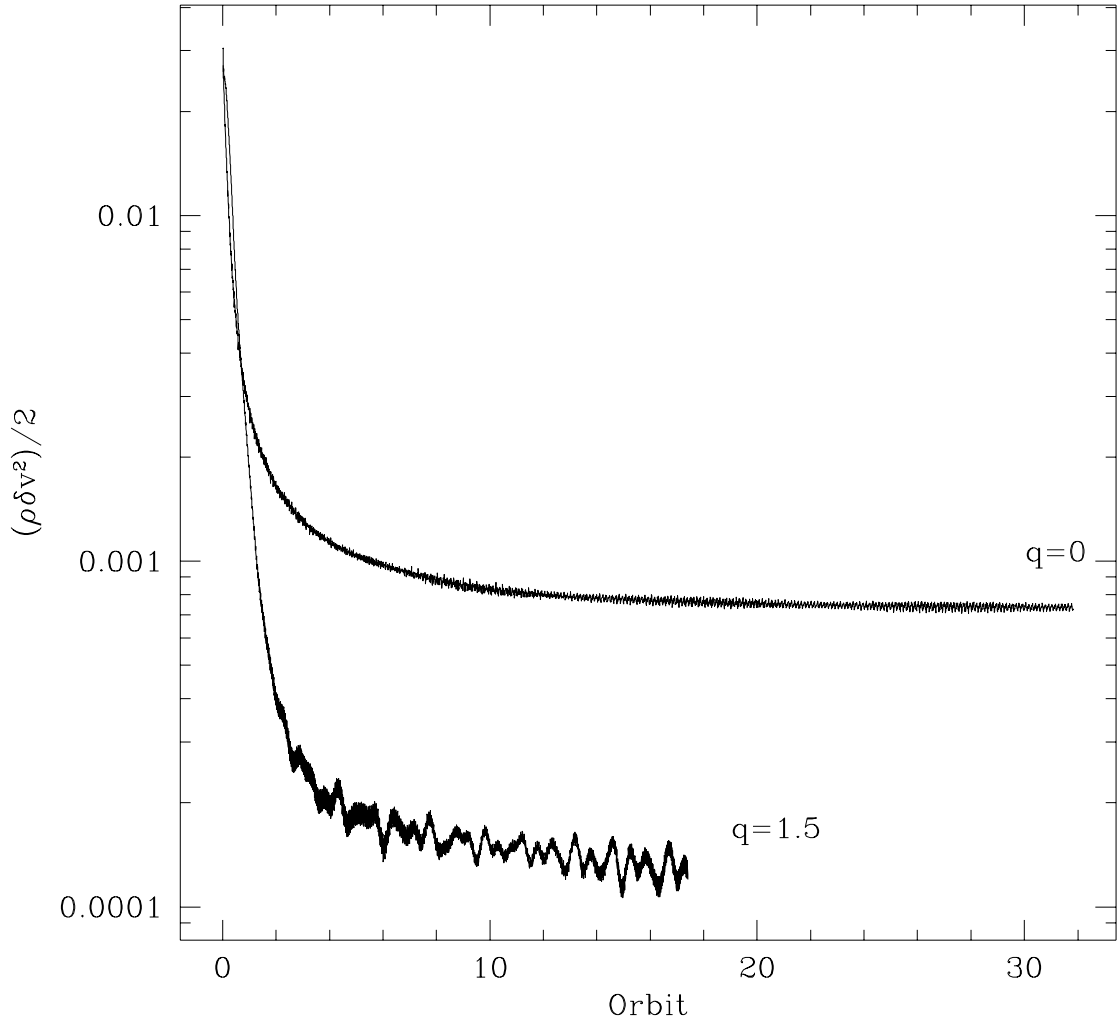}
\caption{
Time evolution of the perturbed kinetic energy in a
constant $\Omega$ simulation, labeled $q=0$, and a 
Keplerian simulation, labeled $q=1.5$.
}
\end{figure}

\begin{figure}
\plotone{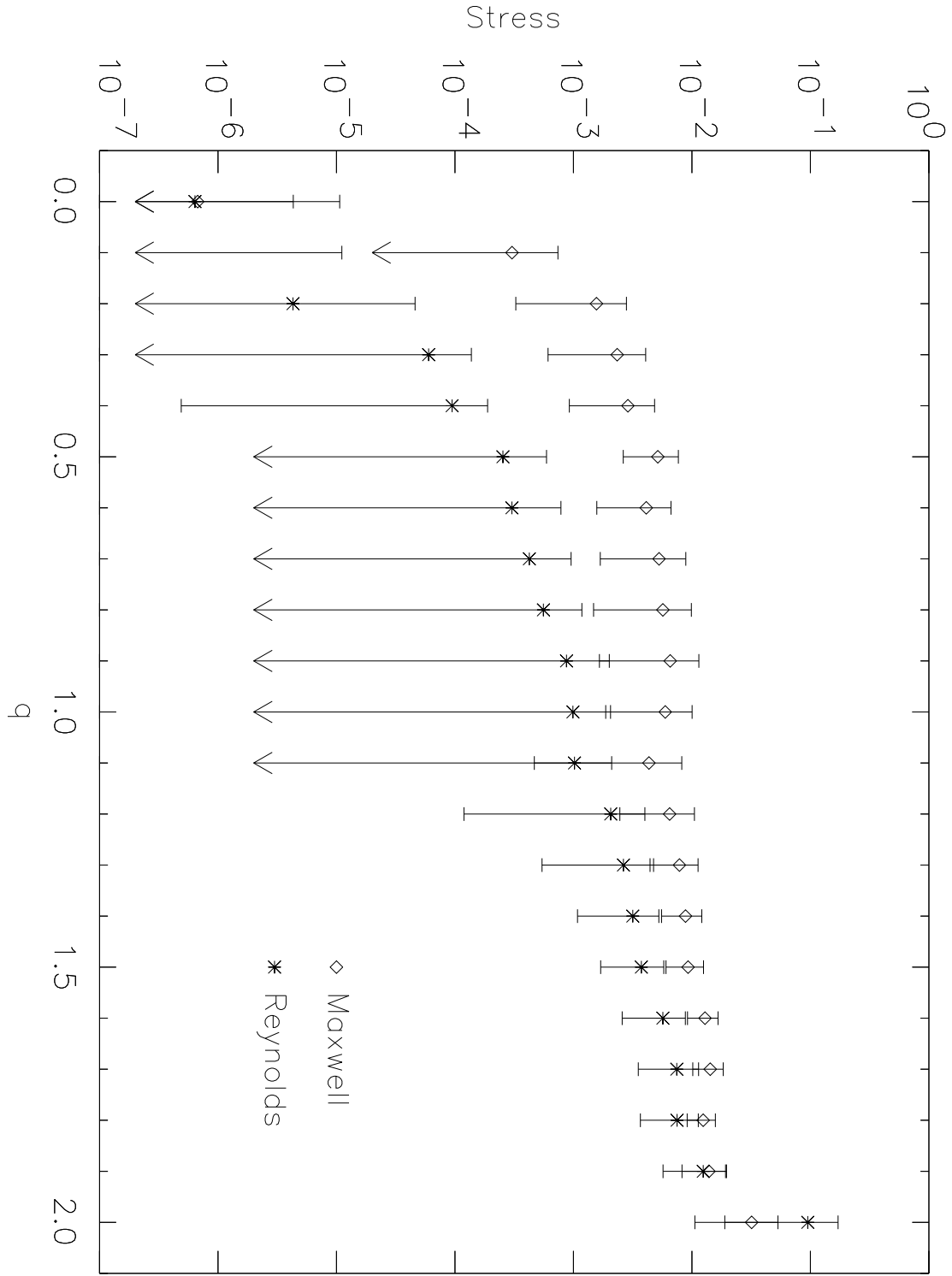}
\caption{
Reynolds stress (stars) and Maxwell stress (diamonds) for a
series of MHD shearing box simulations with different background
angular velocity distributions $q$.  Stress values are time-averaged
over the entire simulation.  Error bars correspond to one standard
deviation in the stress values.
}
\end{figure}

\end{document}